\documentclass[aps,prapplied,showpacs,twocolumn,preprintnumbers,amsmath,amssymb,longbibliography]{revtex4-2}

\usepackage[ruled,lined,boxed,commentsnumbered]{algorithm2e}
\usepackage{amsbsy}
\usepackage{amsfonts}
\usepackage{amssymb}
\usepackage{bm}

\usepackage{graphicx}
\usepackage{dcolumn}
\usepackage{bm}
\usepackage{color}
\usepackage{amsmath}
\usepackage{upgreek}
\usepackage{float}
\usepackage[colorlinks,linkcolor=blue, urlcolor=blue, anchorcolor=blue, citecolor=blue]{hyperref}

\begin{document}

\title{Wavelength-division multiplexing optical Ising simulator enabling fully programmable spin couplings and external magnetic fields }

\author{Li Luo}
\affiliation{Interdisciplinary Center of Quantum Information, State Key Laboratory of Modern Optical Instrumentation, and Zhejiang Province Key Laboratory of Quantum Technology and Device, School of Physics, Zhejiang University, Hangzhou 310027, China}

\author{Zhiyi Mi}
\affiliation{Interdisciplinary Center of Quantum Information, State Key Laboratory of Modern Optical Instrumentation, and Zhejiang Province Key Laboratory of Quantum Technology and Device, School of Physics, Zhejiang University, Hangzhou 310027, China}

\author{Junyi Huang}
\affiliation{Interdisciplinary Center of Quantum Information, State Key Laboratory of Modern Optical Instrumentation, and Zhejiang Province Key Laboratory of Quantum Technology and Device, School of Physics, Zhejiang University, Hangzhou 310027, China}

\author{Zhichao Ruan}
\email{zhichao@zju.edu.cn}
\affiliation{Interdisciplinary Center of Quantum Information, State Key Laboratory of Modern Optical Instrumentation, and Zhejiang Province Key Laboratory of Quantum Technology and Device, School of Physics, Zhejiang University, Hangzhou 310027, China}

\begin{abstract}
Recently, spatial photonic Ising machines (SPIMs) have demonstrated the abilities to compute the Ising Hamiltonian of large-scale spin systems, with the advantages of ultrafast speed and high power efficiency. However, such optical computations have been limited to specific Ising models with fully connected couplings. Here we develop a wavelength-division multiplexing SPIM to enable programmable spin couplings and external magnetic fields as well for general Ising models. We experimentally demonstrate such a wavelength-division multiplexing SPIM with a single spatial light modulator, where the gauge transformation is implemented to eliminate the impact of pixel alignment. To show the programmable capability of general spin coupling interactions, we explore three spin systems: $\pm J$ models, Sherrington-Kirkpatrick models, and only locally connected ${{J}_{1}}\texttt{-}{{J}_{2}}$ models and observe the phase transitions among the spin-glass, the ferromagnetic, the paramagnetic and the stripe-antiferromagnetic phases. These results show that the wavelength-division multiplexing approach has great programmable flexibility of spin couplings and external magnetic fields, which provides the opportunities to solve general combinatorial optimization problems with large-scale and on-demand SPIM.
\end{abstract}
\maketitle

Ising model is an archetypal model widely used for investigations of complex dynamics in physics, computer science, biology, and even social systems. Owing to Moore's law for conventional computers, there has been tremendous interest and a boost in the development of unconventional computing architectures for simulating Ising Hamiltonians, for example, based on optical parametric oscillators \cite{mcmahon2016fully,inagaki2016coherent,inagaki2016large,bohm2019poor,marandi2014network}, lasers \cite{utsunomiya2011mapping, babaeian2019single, tradonsky2019rapid,parto2020realizing, honari2020mapping}, polariton \cite{kalinin2020polaritonic, berloff2017realizing, kalinin2018simulating},  trapped ions \cite{kim2010quantum}, atomic and photonic condensates \cite{struck2013engineering,kassenberg2020controllable}, electronic memorisers \cite{cai2020power}, superconducting qubits \cite{johnson2011quantum, boixo2014evidence, king2018observation}, and nanophotonics circuits  \cite{roques2020heuristic,prabhu2020accelerating,shen2017deep,wu2014optical, okawachi2020demonstration,prabhu2020accelerating}. Despite different approaches and technologies, it is worth noting that the error probability and time-to-solution metrics of these Ising machines have similar scaling trend as a function of the number of spins \cite{mohseni2022ising}. Practically, the difficulty to implement the spin coupling interactions with the proposed hardware has become the main factor limiting scalability and performance for unconventional Ising simulators \cite{heim2015quantum, hamerly2019experimental}. Also complete characterization of possible stable phases is necessary for estimating the Ising description and plays a key role to understand the working principle in spin systems \cite{mertens1998phase, wang2013coherent, strinati2019theory, nishimori2001statistical,inaba2023thermodynamic}.

In this regard, by encoding spins on the phase terms of a monochromatic field with spatial light modulators (SLMs), spatial photonic Ising machines (SPIMs) benefit with reliable large-scale Ising spin systems, even up to thousands of spins by exploring the spatial degrees of freedom  \cite{kumar2020large,pierangeli2021scalable,pierangeli2020adiabatic,pierangeli2020noise, pierangeli2019large, leonetti2021optical,fang2021experimental,huang2021antiferromagnetic, sun2022quadrature,jacucci2022tunable,kumar2023observation}. Like other optical analog computations \cite{silva2014performing,bykov2014optical,ruan2015spatial, youssefi2016analog,zhu2017plasmonic,zhang2018implementing,guo2018photonic,zhu2019generalized,zangeneh2020analogue,zhu2020optical,zhu2021topological}, the calculation of spin system energy is just by instantaneously measuring the light intensity, therefore with ultrafast speed and high power efficiency \cite{pierangeli2019large}. Moreover, to eliminate the impact of pixel alignment, the gauge transformation is proposed to simultaneously encode spin configurations and interaction strengths with a single spatial phase modulator \cite{fang2021experimental}. However, the original proposed SPIM \cite{pierangeli2019large} is only applicable to Mattis-type coupling interactions \cite{mattis1976solvable}. Even with scattering medium, tunable SPIM was demonstrated based on multiple light scattering, while the Ising spin system is still limited to specific fully connected couplings \cite{jacucci2022tunable}. Therefore, how to realize completely programmable spin couplings is a primary target to develop SPIMs for general Ising models.

Here we report a wavelength-division multiplexing SPIM to enable fully programmable spin couplings and external magnetic fields as well. Beyond Mattis-type interaction, we propose a gauge transformation for general Ising models with arbitrary spin interactions. More importantly, with wavelength-division multiplexing, we optically compute the general spin Hamiltonian with the advantages of large scale and ultrafast speed. We experimentally demonstrate such a wavelength-division multiplexing SPIM with a single SLM, where the gauge transformation is implemented to eliminate the impact of pixel alignment. To show the programmable capability of general spin coupling interactions, we explore three kinds of Ising models: fully connected $\pm J$ model, Sherrington-Kirkpatrick (SK) model, and only locally connected ${{J}_{1}}\texttt{-}{{J}_{2}}$ model. For $\pm J$ and SK models, we investigate the process of phase transition with different spin interactions and experimentally observe the spin-glass (SG), ferromagnetic (FM), and paramagnetic (PM) phases by simulating the equilibrium systems at different temperatures. We verify that the critical temperatures evaluated by the wavelength-division multiplexing SPIM are consistent with the predictions of the mean-field theory. We also demonstrate the phase transition from PM to stripe-antiferromagnetic phase in the ${{J}_{1}}\texttt{-}{{J}_{2}}$ model with competing interactions and experimentally observe the presence of SG phase by increasing the stochasticity of the next-nearest-neighbor interactions. These results show the great programmable flexibility of spin couplings and external magnetic fields by the wavelength-division multiplexing SPIM, which provides important potential applications in solving combinatorial optimization problems.

\begin{figure}
\centerline{\includegraphics[width=3.2in] {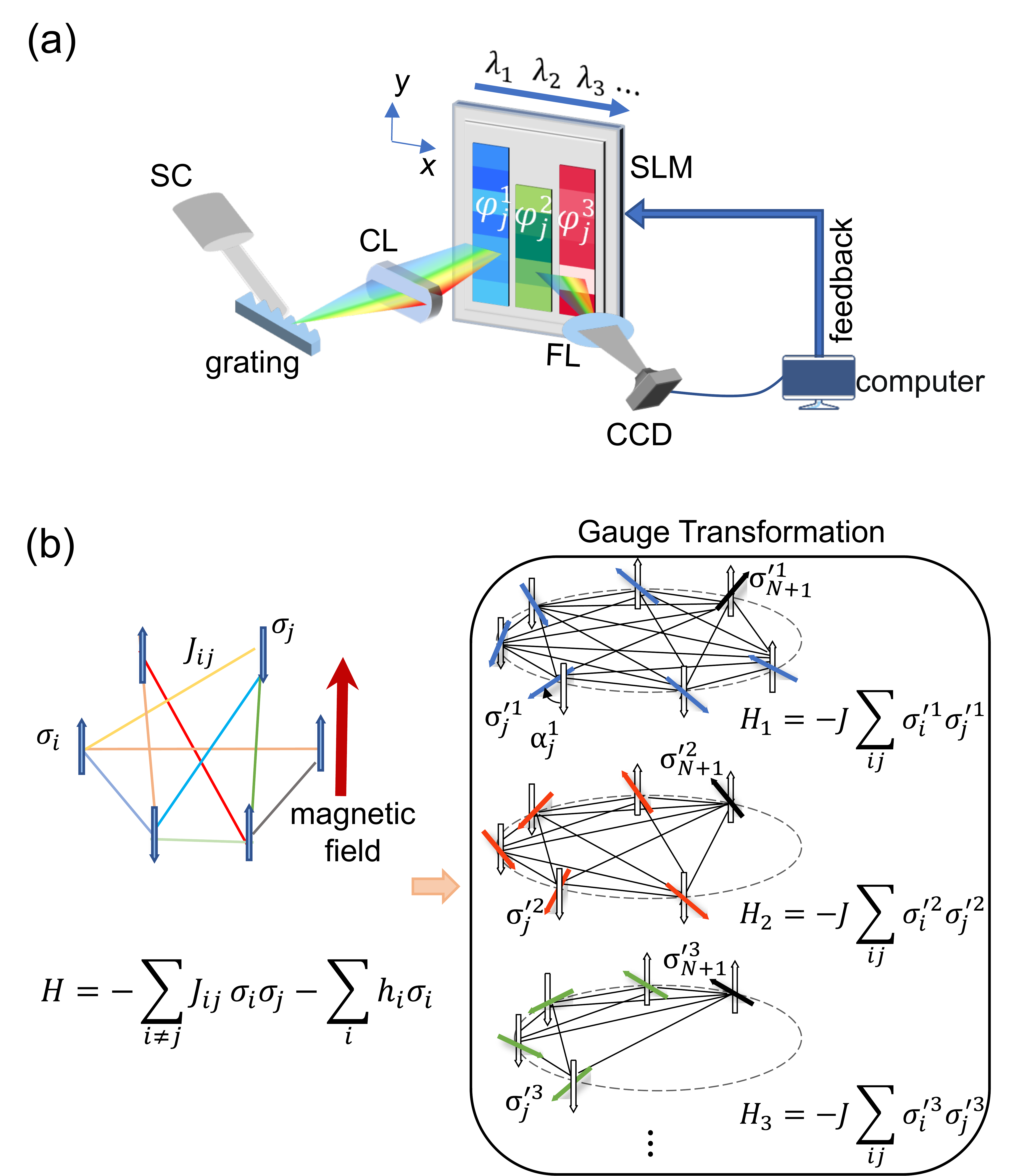}}
\caption{\label{fig:1} (a) Schematic of the wavelength-division multiplexing SPIM. In this setup, light with different wavelengths is diffracted and focus on a phase-only spatial light modulator (SLM) along the $x$-direction, while the pixels in the $y$-direction are coherently illuminated by incident light of the same wavelength. The spins are encoded with phase modulation on the SLM using Eq.~(\ref{eq:2}). SC: super-continuum laser; CL: cylindrical lens; FL: Fourier lens; CCD: charge coupled device camera. (b) The Ising model with general interactions and external magnetic fields is transformed into $N$ numbers of Mattis models via gauge transformation, and $H=\sum_{k=1}^{N}{{{H}_{k}}}+{{H}_{0}}$, where ${{H}_{k}}=-J\sum _{i,j=k}^{N+1}{\sigma_{i}^{\prime k}\sigma_{j}^{\prime k}}$ and $J$ and ${{H}_{0}}$ are constants. Here $\sigma_{j}^{\prime k}$ represents the $z$ component of spin $\sigma_{j}$ rotated by an angle ${\alpha}_{j}^{k}$ with respect to the $z$ axis. }
\end{figure}

The wavelength-division multiplexing SPIM, designed for full programmability of spin interactions and magnetic fields, is shown in Fig.~\ref{fig:1}(a). The illumination components consist of a collimated super-continuum laser, a diffraction grating, and a cylindrical lens. The diffraction grating and cylindrical lens diffract light with different wavelengths onto a SLM along the $x$-axis, while the $y$-axis pixels are illuminated coherently by the same wavelength. By adjusting the diffraction angle, the operation wavelengths are selected within the quasi-stationary region of super-continuum light \cite{genty2010second}. The optical field modulated by the SLM is then transformed by a Fourier lens, resulting in an incoherent intensity summation of different wavelengths and coherent interference for each wavelength at the back focus plane.

Next we show that the wavelength-division multiplexing SPIM can optically compute the Hamiltonian of any Ising model by detecting the intensity at the center position of the back focus plane. The Ising model is given by $H = - \sum\limits_{i \ne j} {{J_{ij}}{\sigma _i}{\sigma _j}}  - \sum\limits_i^N {{h_i}} {\sigma _i}$, where ${J_{ij}}$ and $ {{h}_{i}} $ are the interaction strength and the external magnetic field strength for $N$ spins, respectively, and ${{\sigma }_{i}}$ can take binary values $+1$ or $-1$. By adding an auxiliary spin with ${{\sigma }_{N+1}}=1$, we use a Cholesky-like decomposition to transform the Hamiltonian into multiple Mattis models ${{H}_{k}}=-\sum _{i,j=k}^{N+1}{J\kappa _{i}^{k}\kappa _{j}^{k}{{\sigma }_{i}}{{\sigma }_{j}}}$, so that $H=\sum _{k=1}^{N}{{{H}_{k}}}+{{H}_{0}}$. The interaction strength for the $k$-th Mattis model is given by $J_{ij}^{k}=J\kappa _{i}^{k}\kappa _{j}^{k}$, where $\left| \kappa _{i}^{k} \right|\le 1$, $J$ and ${{H}_{0}}$ are constants with the unit of energy. The transformation is summarized in Sec.~I of the Supplemental Material (SM), and we note that despite the transformation, the degrees of freedom of $\kappa_{i}^k$ remain equal to $N(N+1)/2$, which is the summation of those of $ \{J_{ij}\} $ and $\{h_i\}$ for arbitrary interactions and magnetic fields.

\begin{figure*}
\centerline{\includegraphics[width=5.0in]{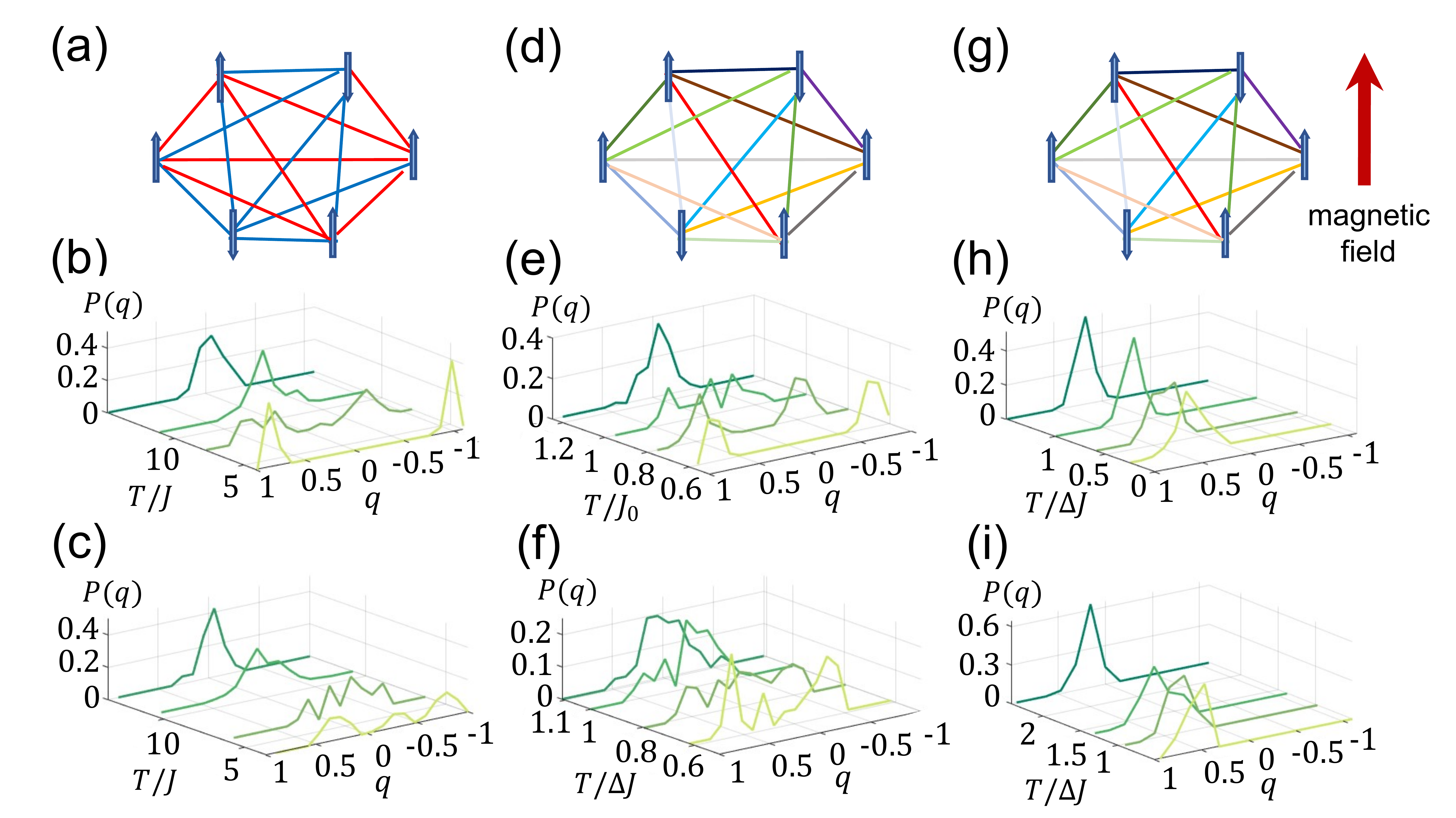}}
\caption{\label{fig:2} Probability distribution of the Parisi parameter ${{q}}$ as a function of $T$, for $N=80$ spins. (a) Schematic representation for the $\pm J$ model. (b) and (c) present the experimental results for $p=0.7$ and $ 0.55$, respectively. (d) Schematic for the SK model. (e) and (f) display the results for ${{J}_{0}}=40$ and $8$, respectively, with $\Delta J=\sqrt{80}$ fixed. (g) Schematic for the SK model with a uniform external magnetic field. (h) and (i) present the results for $h=0.2$ and $2$, respectively, with ${{J}_{0}}$ and $\Delta J$ being the same as (f).}
\end{figure*}

Inspired by the encoding scheme for Mattis model \cite{fang2021experimental, huang2021antiferromagnetic}, here we propose a general one for arbitrary Ising models by a gauge transformation. This transformation allows to encode the spin configurations and program the interaction strengths on a single phase-only SLM, as shown in Fig.~\ref{fig:1}(b). The transformation rotates each original spin about the $z$-axis by an angle ${\alpha }_{j}^{k}=\arccos\kappa_{j}^{k}$ to arrive at a new spin vector, which is then projected onto the $z$-axis to obtain the effective spin $\sigma_{j}^{\prime k}=\kappa_{j}^{k} \sigma_{j}$. By the gauge transformation given as
\begin{equation}
\sigma_{j} \rightarrow \sigma_{j}^{\prime k}, \quad J_{i j}^{k} \rightarrow J,
\end{equation}
the transformed Hamiltonian remains invariant, $H = -J\sum_{k=1}^N{\sum _{i,j=k}^{N+1}{\sigma_i^{'k}\sigma_j^{'k}}} + {H_0}$, where the interactions of the transformed spins are uniform with a strength of $J$ in both short and long ranges.

We encode the gauge-transformed spin configurations $\sigma_{j}^{'k}$ on a single phase-only SLM with wavelength-division multiplexing. For the $k$-th Mattis model, we assume a uniform spectrum intensity light illuminating on SLM and apply a phase modulation on the $y$-directional pixels for each spin as
\begin{equation}
\varphi _{j}^{k}={{\sigma }_{j}}\frac{\pi }{2}+{{(-1)}^{j}}\alpha_{j}^{k}.  \label{eq:2}
\end{equation}
We note that for each Mattis model, the phase modulation should account for the calibration of different wavelengths when encoding $\varphi _{j}^{k}$ in the $x$ direction on the SLM. The normalized intensity $\tilde{I}$ at the center position of the back focus plane is the summation of the incoherent field intensities for light with different wavelengths, $\tilde{I}=\sum_{k=1}^{N}{\sum _{i,j=k}^{N+1}{\sigma_{i}^{'k}\sigma_{j}^{'k}}}$. The Ising Hamiltonian for the general spin interactions is thus optically computed as $H=-J\tilde{I}+{{H}_{0}}$. The details of the gauge transformation and the encoding for general illumination cases are described in SM Sec.~II.

To evaluate the performance of the wavelength-division multiplexing SPIM, we conduct an experiment (detailed experiment setup can be found in SM Sec.III) and simulate three well-studied spin systems: the $\pm J$ model [Fig.~\ref{fig:2}(a)], the SK model [Fig.~\ref{fig:2}(d)] and the SK model under external magnetic field [Fig.~\ref{fig:2}(g)]. The interactions between spins in the $\pm J$ model are either $1$ with a probability of $p$ or $-1$ with a probability of $1-p$. The probability distribution is given as $P\left( {{J}_{ij}} \right)=p\delta \left( {{J}_{ij}}-J \right)+(1-p)\delta \left( {{J}_{ij}}+J \right)$, where $\delta (x)$ is a Kronecker delta function equal to 1 for $x=0$ and 0 for other cases. In the SK model, the probability distribution of ${{J}_{ij}}$ is Gaussian, with the distribution given as $P\left( {{J}_{ij}} \right)=\mathcal{N}({{J}_{ij}};\frac{{{J}_{0}}}{N},\frac{{{{\Delta J}}^2}}{N})$, where ${J}_{0}$ and ${\Delta J}$ are two constant parameters, such that the energy can be extensive and proportional to $N$ \cite{nishimori2001statistical}.

We use the wavelength-division multiplexing SPIM to examine the phase transition with 80 spins. For each model, a quenched realization is generated for the interactions ${{J}_{ij}}$ randomly assigned based on their respective probability distributions. By varying the temperature and utilizing the Markov chain Monte Carlo algorithm, we generate 100 replicas from the final configuration, where each replica is obtained by randomly initializing spin configurations and through 800 iterations of the optical Metropolis Hasting sampling procedure \cite{fang2021experimental,metropolis1953equation, leonetti2021optical}. The spin overlap is then calculated as ${{q}_{\alpha \beta}}=\frac{1}{N}\sum_{i=1}^{N}{\sigma_{i}^{\alpha}}\sigma_{i}^{\beta}$, which measures the similarity between replicas $\alpha$ and $\beta$. The phase transition is characterized by the probability density function $P({q})$ of the overlap.

Figures~\ref{fig:2}(b) and (c) present the experimental results for the $\pm J$ model with the parameters $p=0.7$ and $0.55$, respectively. At high temperatures, both of the two parameters $p=0.7$ and $p=0.55$ result in randomly arranged spins and little correlation between replicas, indicating the PM phase, where $P({{q}})$ has a peak at around zero. At low temperatures, however, $P({{q}})$ displays the distinct density distributions for the two values of $p$. For $p=0.7$, since the interactions are mostly positive, the spins attract each other and the replicas remain in only two ground states of the FM phase. Consequently, $P({{q}})$ at low temperature has two peaks around $1$ and $-1$ as shown in Fig.~\ref{fig:2}(b). On the other hand, for $p=0.55$, the interactions are composed of both positive and negative values, causing frustration during the energy minimization process at low temperatures. This frustration results in a multi-valley energy landscape and $P({{q}})$ takes on a wide range of values at low temperature, as depicted in Fig.~\ref{fig:2}(c). This feature of widely distributed $ {{q}}$ is the hallmark of the SG phase. These results demonstrate that the SG phase transition indeed emerges in the systems simulated by the wavelength-division multiplexing SPIM.

To further study the phase transition with the wavelength-division multiplexing SPIM, we examine the SK model without magnetic fields and compare the estimated critical temperature with the mean-field theory. Fig.~\ref{fig:2}(e) and (f) show the results for ${{J}_{0}}=40$ and $8$, respectively, with $\Delta J=\sqrt{80}$. At high temperatures, both figures show the PM phase with $P({{q}})$ dominated around zero, due to the randomly arranged spin configurations. However, as shown in Figs.~ \ref{fig:2}(e) and (f), $P({{q}})$ at low temperatures exhibit the FM and SG phases for the different values of ${{J}_{0}}$ and occur around $T_c={{J}_{0}}$ and $T_c=\Delta J$, respectively. The experiment results are consistent with the mean-field theory \cite{nishimori2001statistical} with the critical temperatures for the PM-FM transition and for the PM-SG transition.

Additionally, we also investigate the stability of the complex multi-valley energy landscape of the SG phase in the presence of a uniform external magnetic field in the SK model [Fig.~\ref{fig:2}(g)]. As shown in Fig.~\ref{fig:2}(h), for the SK model with ${{J}_{0}}=8$ and $\Delta J=\sqrt{80}$, a weak magnetic field with $h=0.2$ aligns spins with the field direction even at high temperature, causing most ${{q}}$ values to cluster around positive values. Upon decreasing the temperature, the probability distribution of ${{q}}$ resembles that of the SG phase covering a wide range, indicating that weak magnetic fields are not strong enough to completely alter the multi-valley energy landscape. However, when subjected to a stronger magnetic field with $h=2$ [Fig.~\ref{fig:2}(i)], at low temperatures ${{q}}$ are more clustered, which suggests that the magnetic field flattens out some valleys in the energy landscape.

\begin{figure}
\centerline{\includegraphics[width=3.2in]{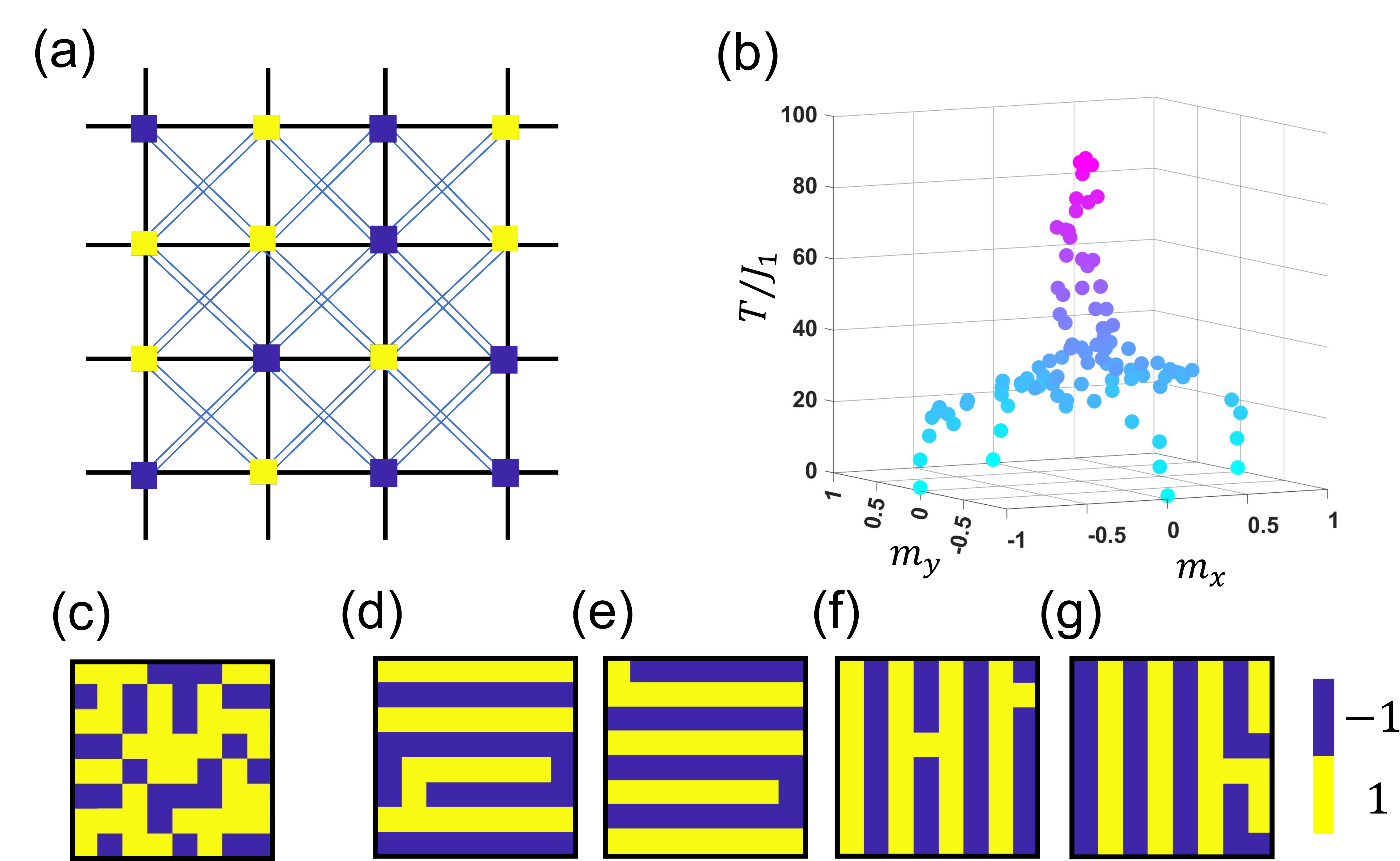}}
\caption{\label{fig:3} Experimental results for a locally connected ${{J}_{1}}\texttt{-}{{J}_{2}}$  model with cyclic boundary condition. (a) A schematic of the ${{J}_{1}}\texttt{-}{{J}_{2}}$ model, where  the thick solid line indicates the nearest neighbor ferromagnetic interaction (${{J}_{1}}>0$) and the blue double line represents the next-nearest-neighbor antiferromagnetic interaction (${{J}_{2}}<0$). Blue and yellow squares denote spins ${\sigma}_{i}=-1$ and $1$, respectively, and the experiment was conducted on an $8\times8$ lattice. (b) Results for the order parameter $(m_x,m_y)$ as a function of $T$, for the parameters ${{J}_{1}}=0.2$ and ${{J}_{2}}=-1$. (c) A spin configuration sampled at $T=70J_1$ representing a PM state. (d-g) Four spin configurations sampled at $T=14.39J_1$ which are adjacent to four striped states, respectively.}
\end{figure}

The wavelength-division multiplexing SPIM can be utilized to study Ising systems beyond the full-coupling models. To demonstrate its full programmability of spin couplings, we examine a locally connected ${{J}_{1}}\texttt{-}{{J}_{2}}$ model [Fig.\ref{fig:3}(a)]. The Hamiltonian only includes nearest-neighbor ferromagnetic  and next-nearest-neighbor antiferromagnetic  interactions on a square lattice with cyclic boundary conditions: $H=-{{J}_{1}}\sum\limits_{\langle ij\rangle }{{{\sigma}_{i}}}{{\sigma }_{j}}-{{J}_{2}}\sum\limits_{\langle \langle ij\rangle \rangle }{{{\sigma }_{i}}}{{\sigma }_{j}}$, where $\langle\rangle$ and $\langle\langle\rangle \rangle$ denote the nearest and next-nearest neighbors, respectively. The nearest-neighbor ferromagnetic interactions align adjacent spins with $J_1>0$ (represented by solid lines in Fig.\ref{fig:3}(a)), while the next-nearest-neighbor antiferromagnetic interactions drive two adjacent rows and columns to have opposite orientations with $J_2<0$ (represented by double parallel lines).  When the antiferromagnetic interaction $J_2$ is strong enough to overcome the ferromagnetic interaction $J_1$, a striped phase is produced, characterized by a two-component order parameter $(m_x,m_y)$, where $m_{x}=\frac{1}{N} \sum_{i=1}^{N} \sigma_{i}(-1)^{x_{i}}$, and ${{m}_{y}}=\frac{1}{N} \sum_{i=1}^{N} {{{\sigma}_{i}}}{{(-1)}^{{{y}_{i}}}}$, and $(x_i,y_i)$ are the coordinates of spin $\sigma_i$. The ground states of the ${{J}_{1}}\texttt{-}{{J}_{2}}$ model are ${{\mathbb{Z}}_{4}}$ ordered when $|J_2|/J_1>1/2$, and can be represented by the order parameters of $(m_x, m_y)=(\pm1, 0)$ and $(m_x, m_y)=(0, \pm1)$, corresponding to two longitudinal and transverse striped states with opposite spin directions, respectively \cite{jin2013phase}.

Using the wavelength-division multiplexing SPIM, we simulate the transition between the striped states and the PM phases. Fig.~\ref{fig:3}(b) shows the simulation results for the parameters ${{J}_{1}}=0.2$ and ${{J}_{2}}=-1$ on an $8\times8$ lattice. Above the critical temperature of $T_c=20J_1$, the order parameters $(m_x,m_y)$ are both close to zero due to the thermal fluctuations destroying the long-range correlation between spins as a sample of spin configurations shown in Fig.~\ref{fig:3}(c). However, below $T_c$, the order parameters split into four clusters located in different quadrants, corresponding to the four striped states [Fig.~\ref{fig:3}(b)]. Figs.~\ref{fig:3}(d-g) show four samples of the spin configurations, which exhibit a long stripe spatial distribution. These results for the ${{J}_{1}}\texttt{-}{{J}_{2}}$ model demonstrate the programmability of spin couplings.

\begin{figure}
\centerline{\includegraphics[width=3.4in]{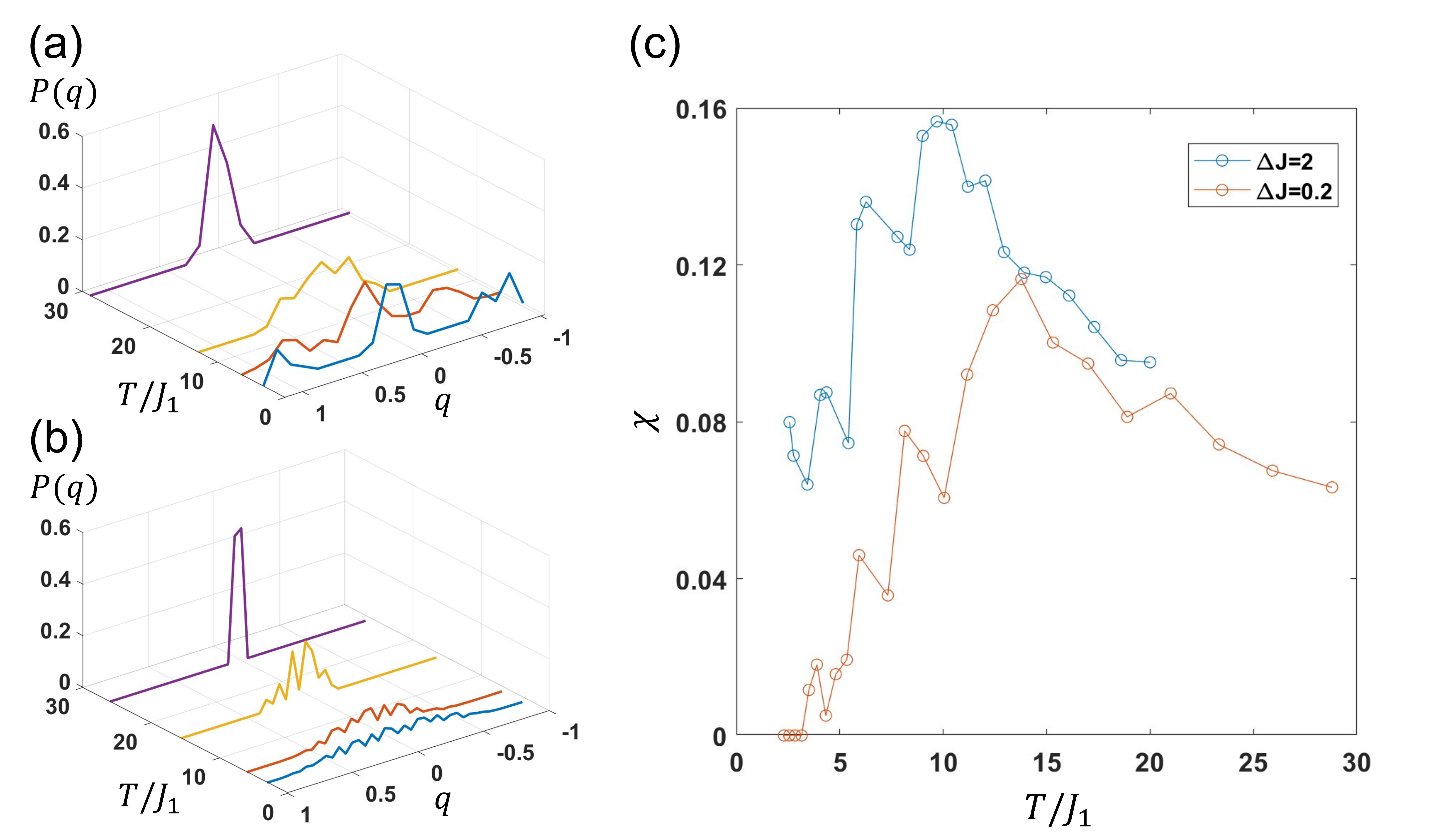}}
\caption{\label{fig:4} (a) and (b) Probability density distribution of ${{q}}$ for the ${{J}_{1}}\texttt{-}{{J}_{2}}$ models with the fixed nearest neighbor interactions ${{J}_{1}}=1$, while the next-nearest-neighbor interactions are Gaussian distributed with the mean value of ${{J}_{0}}=-1$ and the standard deviation of $\Delta J=0.2$ and $2$, respectively. (c) Experimental measured susceptibility for $\Delta J=0.2$ and $\Delta J=2$.}
\end{figure}

The ${{J}_{1}}\texttt{-}{{J}_{2}}$ model plays a crucial role in comprehending the low-temperature phase of short-range spin glass. With the help of the wavelength-division multiplexing SPIM, we consider the next-nearest-neighbor interactions are Gaussian as  $P\left( {{J}_{2}} \right)= \mathcal{N}({{J}_{2}}; {{J}_{0}},{\Delta J}^2)$ with a mean value of ${{J}_{0}}=-1$ and a standard deviation $\Delta J$, while maintaining a fixed value of nearest-neighbor interactions ${{J}_{1}}=1$. For small variances of $\Delta J=0.2$ [Fig.~\ref{fig:4}(a)], the density distribution of ${{q}}$ exhibits a single peak at high temperatures, signifying the PM phase. However, as the temperature decreases, $P({{q}})$ transforms into three clusters of peaks, indicative of the ${{\mathbb{Z}}_{4}}$ ordered striped ground states. The peak around ${{q}}=0$ is twice as high as those around ${{q}}=\pm 1$. In contrast, by increasing the standard deviation to $\Delta J=2$, the increased disorder in the next-nearest-neighbor interactions results in a SG phase at low temperatures. The SG phase is characterized by many pairs of ground states and results in multiple sharp peaks of $P({{q}})$ as Fig.\ref{fig:4}(b).

We also experimentally measure the susceptibility $\chi = \frac{1}{N}\sum_{i = 1}^N {\frac{{1 - m_i^2}}{T}}$\cite{fischer1993spin} to investigate the transition temperature in the ${{J}_{1}}\texttt{-}{{J}_{2}}$ model, where ${{m}_{i}}$ represents the time ensemble average of ${{\sigma}_{i}}$.  At high temperatures, thermodynamic fluctuations cause the spin orientation to constantly change, resulting in ${{m}_{i}}$ being close to zero, as shown in Fig.~\ref{fig:4}(c). The susceptibility in both two cases decreases with increasing temperature and scales with $\frac{1}{T}$. The results also indicate that the critical temperatures are different, with ${{T}_{c}}=14{{J}_{1}}$ and ${{T}_{c}}=10{{J}_{1}}$ for $\Delta J=0.2$ and $\Delta J=2$, respectively. In the case of $\Delta J=0.2$, the interactions of the next-nearest-neighbors are relatively small and close to $-1$. As the temperature decreases below ${{T}_{c}}$, the system reaches the striped ground states with ${{m}_{i}}$ approaching $\pm 1$, causing the average susceptibility $\chi $ to converge to zero. In contrast, for $\Delta J=2$, at low temperatures, the local spin orientation still varies slowly due to the presence of numerous ground states in the SG phase, resulting in $\chi$ being close to a non-zero constant. These experimental results clearly demonstrate the spin-glass phase transition in the ${{J}_{1}}\texttt{-}{{J}_{2}}$ model with short-range interactions.

In summary, we propose the wavelength-division multiplexing SPIM to realize fully programmable spin couplings and external magnetic fields. With the full-coupling $\pm J$ models, SK models and the only locally connected ${{J}_{1}}\texttt{-}{{J}_{2}}$ models, we demonstrate the great programmable flexibility of spin couplings and external magnetic fields with the wavelength-division multiplexing SPIM. By simulating the equilibrium systems at different temperatures, we experimentally observe the phase transitions among the spin-glass, the ferromagnetic, the paramagnetic and the stripe-antiferromagnetic phases with the wavelength-division multiplexing SPIM. The exhibited rich phase diagrams are beneficial to searching for the true ground state of the Ising model, and thus the wavelength-division multiplexing approach provides important potential applications in solving combinatorial optimization problems.

The authors acknowledge funding through the National Key Research and Development Program of China (2022YFA1405200) and the National Natural Science Foundation of China (12174340).

%


\clearpage
\newpage

\setcounter{section}{0}

\newcommand{\hbAppendixPrefix}{S}
\renewcommand{\thefigure}{\hbAppendixPrefix\arabic{figure}}
\setcounter{figure}{0}

\renewcommand{\thetable}{\hbAppendixPrefix\arabic{table}}
\setcounter{table}{0}
\renewcommand{\theequation}{\hbAppendixPrefix\arabic{equation}}
\setcounter{equation}{0}
\onecolumngrid


\begin{center}
\large{\textbf{Supplementary Material: Wavelength-division multiplexing optical Ising simulator enabling fully programmable spin couplings and external magnetic fields}}
\end{center}


\begin{center}
Li Luo, Zhiyi Mi, Junyi Huang, and Zhichao Ruan
\end{center}

\begin{center}
\small{\textit{Interdisciplinary Center of Quantum Information, \\
State Key Laboratory of Modern Optical Instrumentation, \\
and Zhejiang Province Key Laboratory of Quantum Technology and Device, \\
 Department of Physics, Zhejiang University, Hangzhou 310027, China}}
\end{center}

\section{Decomposition of general Ising Hamiltonian into multiple Mattis-type models}

We first consider the Hamiltonian of Ising model without external magnetic field as $H=-\sum\limits_{i\ne j}^{N}{{{J}_{ij}}{{\sigma}_{i}}}{{\sigma }_{j}}$, and $N$ is the number of spins. To facilitate decomposition of the interaction matrix $\boldsymbol{J}$, we introduce auxiliary positive diagonal elements ${{J}_{ii}}$, and propose a Cholesky-like algorithm (Algorithm 1). This algorithm expresses $\boldsymbol{J}$ as
\begin{equation}
[{{J}_{ij}}]={{\mathbf{\xi }}^{1}}\times {{\mathbf{(}{{\mathbf{\xi }}^{1}}\mathbf{)}}^{T}}+{{\mathbf{\xi }}^{2}}\times {{\mathbf{(}{{\mathbf{\xi }}^{2}}\mathbf{)}}^{T}}+{{\mathbf{\xi }}^{3}}\times {{\mathbf{(}{{\mathbf{\xi }}^{3}}\mathbf{)}}^{T}}+\cdots +{{\mathbf{\xi }}^{N}}\times {{\mathbf{(}{{\mathbf{\xi }}^{N}}\mathbf{)}}^{T}} \label{eq:S1}
\end{equation}
where ${{\mathbf{\xi }}^{1}}=\left(\begin{matrix}\xi _{1}^{1}\\ \xi _{2}^{1}\\ \xi _{3}^{1}\\ \vdots\\ \xi _{N}^{1}\\ \end{matrix}\right)$, ${{\mathbf{\xi }}^{2}}=\left(\begin{matrix} 0 \\ \xi _{2}^{2} \\ \xi _{3}^{2}\\ \vdots \\ \xi _{N}^{2}\\ \end{matrix} \right)$, ${{\mathbf{\xi }}^{3}}=\left(\begin{matrix} 0\\ 0\\ \xi _{3}^{3}\\ \vdots \\ \xi _{N}^{3}  \\ \end{matrix} \right)$, $\cdots$, ${{\mathbf{\xi }}^{N}}=\left( \begin{matrix} 0 \\ 0 \\ 0 \\ \vdots \\ \xi _{N}^{N} \\ \end{matrix} \right)$. We note that the auxiliary diagonal elements ${{J}_{ii}}$ correspond to the self-interaction of spins, and only shift the energy value, without affecting the spin configurations of the ground state.

\begin{algorithm}
\caption{Cholesky-like decomposition}
\label{alg:algorithm1}
\KwIn{Number of spins $N$, Interaction matrix $\boldsymbol{J}=\{J_{ij}\}$.}
\KwOut{Lower triangular matrix $\boldsymbol{\xi}$, which satisfies $J_{jk}=\sum_{i=1}^N\xi_k^i\xi_k^i$ when $j\neq k$.}
Define $\boldsymbol{\xi}=\begin{pmatrix}\xi_1^1&\xi_1^2&\cdots&\xi_1^N\\\xi_2^1&\xi_2^2&\cdots&\xi_2^N\\\vdots&\vdots&\ddots&\vdots\\\xi_N^1&\xi_N^2&\cdots&\xi_N^N\end{pmatrix}_{N\times N}=\begin{pmatrix}0&0&\cdots&0\\0&0&\cdots&0\\\vdots&\vdots&\ddots&\vdots\\0&0&\cdots&0\end{pmatrix}_{N\times N}$, $\boldsymbol{J}^{now}=\boldsymbol{J}$\;
		
\For{$i=1$ to $N$}
{		
$\xi^i_i=\sqrt{J_{ii}^{now}}$\;
		
\For{$j=i+1$ to $N$}
{
$\xi_j^i=\frac{\boldsymbol{J}^{now}_{ji}}{\xi^i_i}$\;
}
$\boldsymbol{J}^{now}=\boldsymbol{J}^{now}-\boldsymbol{\xi}^{i}\times(\boldsymbol{\xi}^{i})^T$\;
}
\end{algorithm}

We now consider the Ising model with an external magnetic field, given by the Hamiltonian $H=-\sum\limits_{i\ne j}^{N}{{{J}_{ij}}{{\sigma }_{i}}}{{\sigma }_{j}}-\sum\limits_{i}^{N}{{{h}_{i}}}{{\sigma }_{i}}$. To develop a Cholesky-like algorithm for this model, we add auxiliary elements ${{\theta }^{k}}$ to each vector ${\kappa ^{1}} = \left( {\begin{array}{*{20}{c}} {\xi _1^1}\\ {\xi _2^1}\\ {\xi _3^1}\\ \vdots \\ \begin{array}{l} \xi _N^1\\
{\theta ^1} \end{array} \end{array}} \right)$, ${\kappa ^2} = \left({\begin{array}{*{20}{c}} 0\\ {\xi _2^2}\\ {\xi _3^2}\\\vdots \\ \begin{array}{l} \xi _N^2\\ {\theta ^2} \end{array} \end{array}} \right)$, ${\kappa ^3} = \left({\begin{array}{*{20}{c}} 0\\ 0\\ {\xi _3^3}\\  \vdots \\  \begin{array}{l} \xi _N^3\\ {\theta ^3} \end{array} \end{array}} \right)$, $\cdots$, ${\kappa ^N} = \left( {\begin{array}{*{20}{c}} 0\\ 0\\ 0\\  \vdots \\  \begin{array}{l} \xi _N^N\\ {\theta ^N} \end{array} \end{array}} \right)$. The values of ${{\theta }^{k}}$ are derived using Algorithm 2.

We further normalize each vector ${{\kappa }^{k}}$ by dividing it by $\sqrt{J_{}}$, where ${{J}_{{}}}=\max \{|\kappa _{j}^{k}{{|}^{2}}\}$, so that $-1\le \kappa _{j}^{k}\le 1$. Next, we introduce an auxiliary spin $\sigma_{N+1}=1$ and rewrite the Hamiltonian as:
\begin{equation}
    H=-J\sum\limits_{k}^{N}{\sum\limits_{ij}^{N+1}{\kappa_{i}^{k}\kappa_{j}^{k}{{\sigma }_{i}}{{\sigma }_{j}}}}+{{H}_{0}} \label{eq:S2}
\end{equation}
where ${{H}_{0}}=J(\sum\limits_{k}^{N}{{{(\xi _{k}^{k})}^{2}}}+\sum\limits_{k}^{N}{{{(\theta _{{}}^{k})}^{2}})}$.

\begin{algorithm}
		\KwIn{Magnetic field parameter $\boldsymbol{h}=\{h_1,h_2,...,h_N\}$, lower triangular matrix $\boldsymbol{\xi}$.}
		\KwOut{Lower triangular matrix $\boldsymbol{\kappa}$, which satisfies $J_{jk}=\sum_{i=1}^N\kappa_j^i\kappa_k^i$ when $j\neq k$, and $\sum_{i=1}^{N}\kappa_{j}^{i}\kappa_{N+1}^{i}=\frac{h_{j}}{2}$.}
	\caption{Magnetic field encoding method}
	\label{alg:algorithm2}
		Define $\boldsymbol{\theta}=(\theta^1\quad\theta^2\quad\cdots\quad\theta^N)$\;

\For{$i=1$ to $N$}
{
$\theta^i=\frac{\frac{h_i}{2}-\sum^{i-1}_{j=1}\theta^j\xi_i^j}{\xi_i^i}$\;
}
\end{algorithm}

\section{Gauge transformation and Optical computation of general Ising Hamiltonian}

\begin{figure}
\centerline{\includegraphics[width=4.5in]{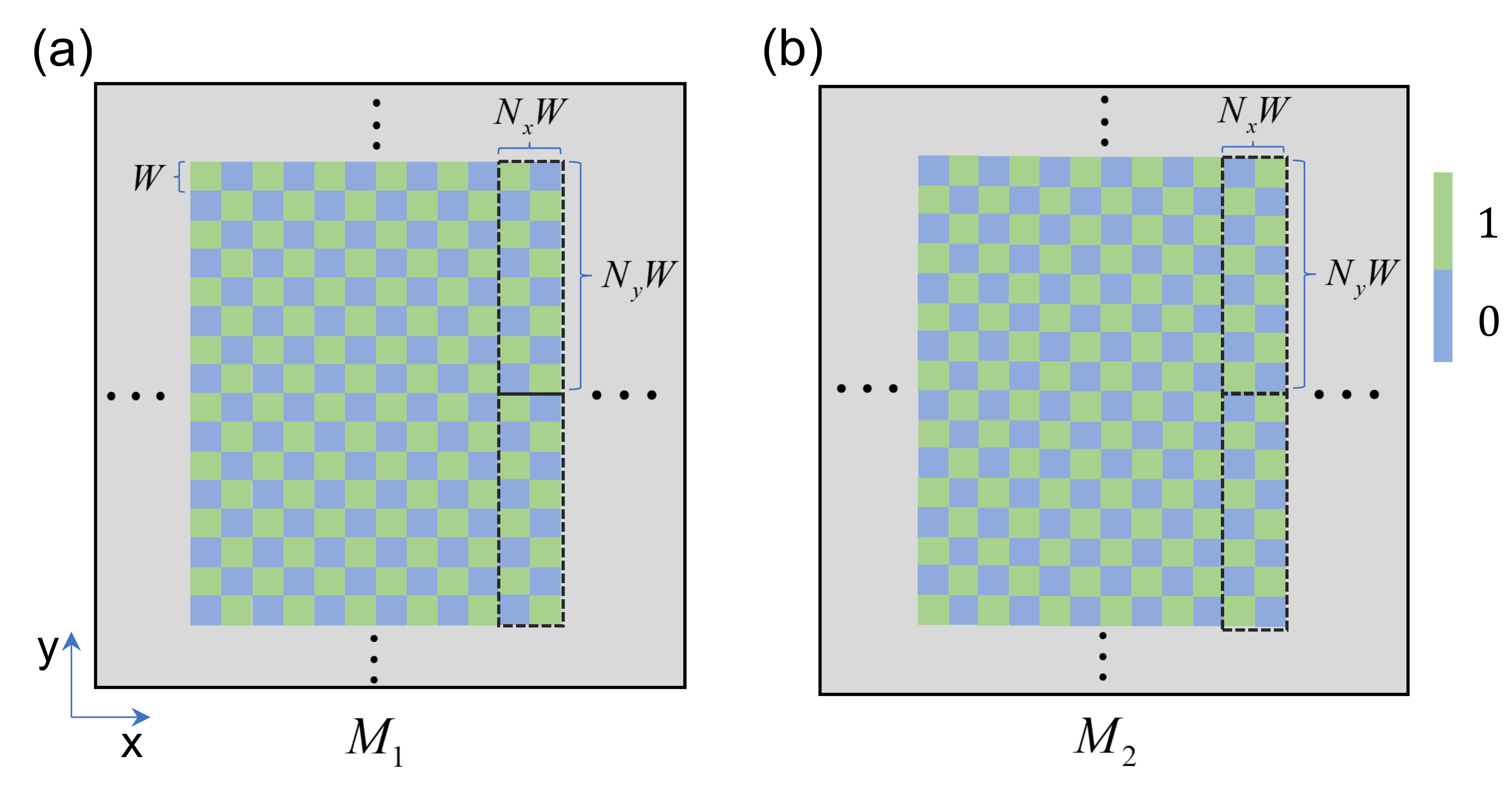}}
\caption{\label{fig:S1}(a) and (b) are the checkerboard functions ${{M}_{1}}$ and ${{M}_{2}}$. Here each spin is encoded by ${{N}_{y}} \times {{N}_{x}}$ pixels as shown in the black dashed box. For the $k$-th Mattis-type model, the center of the $j$-th spin is located at $(k{{N}_{x}}W, j{{N}_{y}}W)$, where $W$ represents the pixel width of the SLM.}
\end{figure}

 We first assume that the amplitude of the illumination light on SLM for each wavelength is uniform and equal to ${A_0}$. For the $k$-th Mattis-type model, with the gauge transformation proposed in the main text, we encode each spin by ${{N}_{y}} \times {{N}_{x}}$ pixels as Fig.~\ref{fig:S1} and the phase modulation on SLM ${\varphi}^k_j$ as defined in Eq.(2) in the main text. Therefore, the electric fields after the SLM modulation are given by
\begin{equation}
{E_k}(x,y)= iA_0({F_+} \cdot {M_1} + {F_-} \cdot {M_2}) \label{eq:S3}
\end{equation}
where $F_+$ and $F_-$ are written as
\begin{equation}
\begin{aligned}
{F_\pm }&=\sum\limits_{j=k}^{N + 1} {{e^{\pm i\alpha _j^k}\sigma _j}{{{\mathop{\rm Rect}\nolimits} }_{{N_y}W}}\left( {y - y_j} \right) \cdot {{{\mathop{\rm Rect}\nolimits} }_{{N_x}W}}(x) } \notag
\end{aligned}
\end{equation}
Here $(x,y)$ denotes the spatial coordinate on the SLM plane, and $y_j=j{{N}_{y}}W$ represents the position of the $j$-th spin in $y$ direction on SLM. The rectangular function is defined as
\begin{equation}
{{\mathop{\rm Rect}\nolimits} _{{N_y}W}}(y) ={\rm{Rect}}(\frac{y}{N_yW})= \left\{ {\begin{array}{*{20}{l}} {1}, &{|y| \le {N_y}W/2}\\ 0, &{|y| > {N_y}W/2} \end{array}} \right.
\end{equation}
where $W$ represents the pixel width of the SLM. $M_1$ and $M_2$ are two checkerboard functions, as shown in Fig.~\ref{fig:S1}, such that ${M_1} + {M_2} = 1 $, and are defined as
\begin{equation}
\begin{array}{c}
{M_1} = \sum\limits_{m,n = -\infty }^\infty {[\delta(x - 2mW,y - 2nW) +\delta(x - (2m + 1)W,y-(2n + 1)W)]} \otimes {{\mathop{\rm Rect}\nolimits} _W}(x,y) \\
{M_2} = \sum\limits_{m,n =  - \infty }^\infty  {[\delta (x - 2mW,y - (2n + 1)W) + \delta (x - (2m + 1)W,y - 2nW)]}  \otimes {{\mathop{\rm Rect}\nolimits} _W}(x,y) \notag
\end{array}
\end{equation}
where $\otimes $ denotes the convolution operation.

After passing through the Fourier lens, the optical fields at the CCD for each wavelength are evaluated using a two-dimensional Fourier transform. The resulting electric fields in the spatial spectra are given by:
\begin{equation}
{\widetilde E_k}({k_x},{k_y}) = \frac{iA_0}{{4}}({G_+} \otimes {P_1} + {G_-} \otimes {P_2}) \notag
\end{equation}
where
\begin{equation}
{G_\pm}=\sum\limits_{j=k}^{N + 1} {{e^{\pm i\alpha _j^k}}} {\sigma _j} \cdot {{\mathop{\rm sinc}\nolimits} _{{N_y}W}}\left( {{k_y}} \right){e^{i{k_y}{y_j}}} \cdot ({N_y}W) \cdot {{\mathop{\rm sinc}\nolimits} _{{N_x}W}}({k_x}) \cdot ({N_x}{\rm{W)}} \notag
\end{equation}
\begin{equation}
{P_1}=\sum\limits_{m,n =  - \infty }^\infty  {\left( {1 + {{( - 1)}^{m + n}}} \right)} \,\delta \left( {{k_x} - m\frac{\pi }{W},{k_y} - n\frac{\pi }{W}} \right) \cdot{{\mathop{\rm sinc}\nolimits} _W}\left( {{k_x},{k_y}} \right) \notag
\end{equation}
\begin{equation}
{P_2}=\sum\limits_{m,n =  - \infty }^\infty  {\left( {{{( - 1)}^m} + {{( - 1)}^n}} \right)} \,\delta \left( {{k_x} - m\frac{\pi }{W},{k_y} - n\frac{\pi }{W}} \right) \cdot {{\mathop{\rm sinc}\nolimits} _W}\left( {{k_x},{k_y}} \right) \notag
\end{equation}
\begin{equation}
{{\mathop{\rm sinc}\nolimits}_W}({k_x},{k_y}) = {\mathop{\rm sinc}\nolimits}(\frac{k_xW}{2\pi}) \cdot {\mathop{\rm sinc}\nolimits}(\frac{k_yW}{2\pi})\notag
\end{equation}
\begin{equation}
{{\mathop{\rm sinc}\nolimits} _{{N_x}W}}({k_x}) ={\mathop{\rm sinc}\nolimits}(\frac{k_xN_xW}{2\pi})  \notag
\end{equation}
\begin{equation}
{{\mathop{\rm sinc}\nolimits} _{{N_y}W}}({k_y}) = {\mathop{\rm sinc}\nolimits}(\frac{k_yN_yW}{2\pi})  \notag
\end{equation}
\begin{equation}
{{\mathop{\rm sinc}\nolimits}(k)} = \frac{{\mathop{\rm sin}\nolimits}(k\pi)}{k\pi} \notag
\end{equation}
Here, the convolution terms indicate that due to the checkerboard modulation, the beams are diffracted at multiple orders around $(m\frac{\pi }{W}, n\frac{\pi }{W})$ in the angular spectral space, where $m$ and $n$ are two integers. Assuming negligible field overlap between different orders, the zeroth order diffraction field for $m=0$ and $n=0$ can be expressed as
\begin{equation}
\begin{aligned}
{{\tilde E}_k}({k_x},{k_y}) \doteq i{A_0}C\sum\limits_{j=k}^{N + 1} {\kappa _j^k{\sigma _j}} {e^{i{k_y}{y_j}}}{{\mathop{\rm sinc}\nolimits} _W}({k_x},{k_y}){{\mathop{\rm sinc}\nolimits} _{{N_x}W}}\left( {{k_x}} \right){{\mathop{\rm sinc}\nolimits} _{{N_y}W}}\left( {{k_y}} \right)
\end{aligned}
\end{equation}
where $C={N}_{x}W\cdot {{N}_{y}}W$.

For the angle spectrum $({{k}_{x}},{{k}_{y}})$, the spatial coordinate $({{u}_{x}},{{u}_{y}})$ on the detection plane are ${{u}_{x}}={{k}_{x}}\frac{f\lambda }{2\pi }$, ${{u}_{y}}={{k}_{y}}\frac{f\lambda }{2\pi }$. Thus, the intensity contributed by the $k$-th wavelength light is given by
\begin{equation}
\begin{aligned}
{I_k}({u_x},{u_y})&= \tilde E_k^*\cdot {{\tilde E}_k}\\ &=A_0^2{C^2}\sum\limits_{i,j=k}^{N + 1} {\kappa _i^k\kappa _j^k{\sigma _i}{\sigma _j}} {e^{i\frac{{2\pi }}{{f\lambda }}{u_y}({y_j} - {y_i})}}{{\mathop{\rm sinc}\nolimits} ^2}\left( {\frac{{u_xW}}{{f\lambda }}} \right){{\mathop{\rm sinc}\nolimits} ^2}\left( {\frac{{u_yW}}{{f\lambda }}} \right){{\mathop{\rm sinc}\nolimits} ^2}\left( {\frac{{{u_x}W{N_x}}}{{f\lambda }}} \right){{\mathop{\rm sinc}\nolimits} ^2}\left( {\frac{{{u_y}W{N_y}}}{{f\lambda }}} \right) \notag
\end{aligned}
\end{equation}
In particular, the intensity at ${{u}_{x}}=0$ and ${{u}_{y}}=0$ is
\begin{equation}
{I_k}(0,0) = A_0^2{C^2}\sum\limits_{i,j=k}^{N + 1} {\kappa _i^k\kappa _j^k{\sigma _i}{\sigma _j}} \label{eq:S6}
\end{equation}

Due to the incoherent nature between different wavelengths, the total intensity detected at the zeroth order is
\begin{equation}
\begin{aligned}
I &= \sum\limits_k^N {{I_k}} (0,0)\\
 &= A_0^2{C^2}\sum\limits_k^N {\sum\limits_{i,j=k}^{N + 1} {\kappa _i^k\kappa _j^k{\sigma _i}{\sigma _j}} }
\end{aligned}
\end{equation}
For a special case of $\{\kappa _j^k=1\}$ and $\{\sigma_j=1\}$ for all spins, the phase modulations are  given by $\{ {\varphi}^k_j=\frac{\pi}{2} \}$ (see Eq.~(2) in the main text). That is, when the phase modulation is set to $\frac{\pi}{2}$ for all pixels, the intensity on CCD is ${{I}_{0}}=A_0^2{C^2}N(N+1)^2$. Therefore by measuring the intensities and normalizing the intensity to ${{I}_{0}}$, the Ising Hamiltonian for the general spin interaction is optically computed as
\begin{equation}
\tilde I={N(N+1)^2} \frac{I}{I_0} = \sum\limits_k^N {\sum\limits_{i,j=k}^{N + 1} {\kappa _i^k\kappa _j^k{\sigma _i}{\sigma _j}} }
\end{equation}

We note that when the amplitude for different wavelengths is not uniform, supposing that the amplitude for the $k$-th wavelength is $A{{}_{k}}$, the phase encoding in Eq.~(2) can be modified as $\alpha _j^k = \arccos \frac{{\kappa _j^k}}{{\frac{{A_k}}{{A_{\min}}}}}$. Here $A_{\min}$ is the minimum amplitude among all wavelengths. Under such circumstances, the above derivation is still available, but the normalized intensity ${{I}_{0}}$ changes to ${I_0} = A_{\min}^2{C^2}N(N+1)^2$.

\section{Experimental setup and measurement of intensity}

\begin{figure}
\centerline{\includegraphics[width=4.0in]{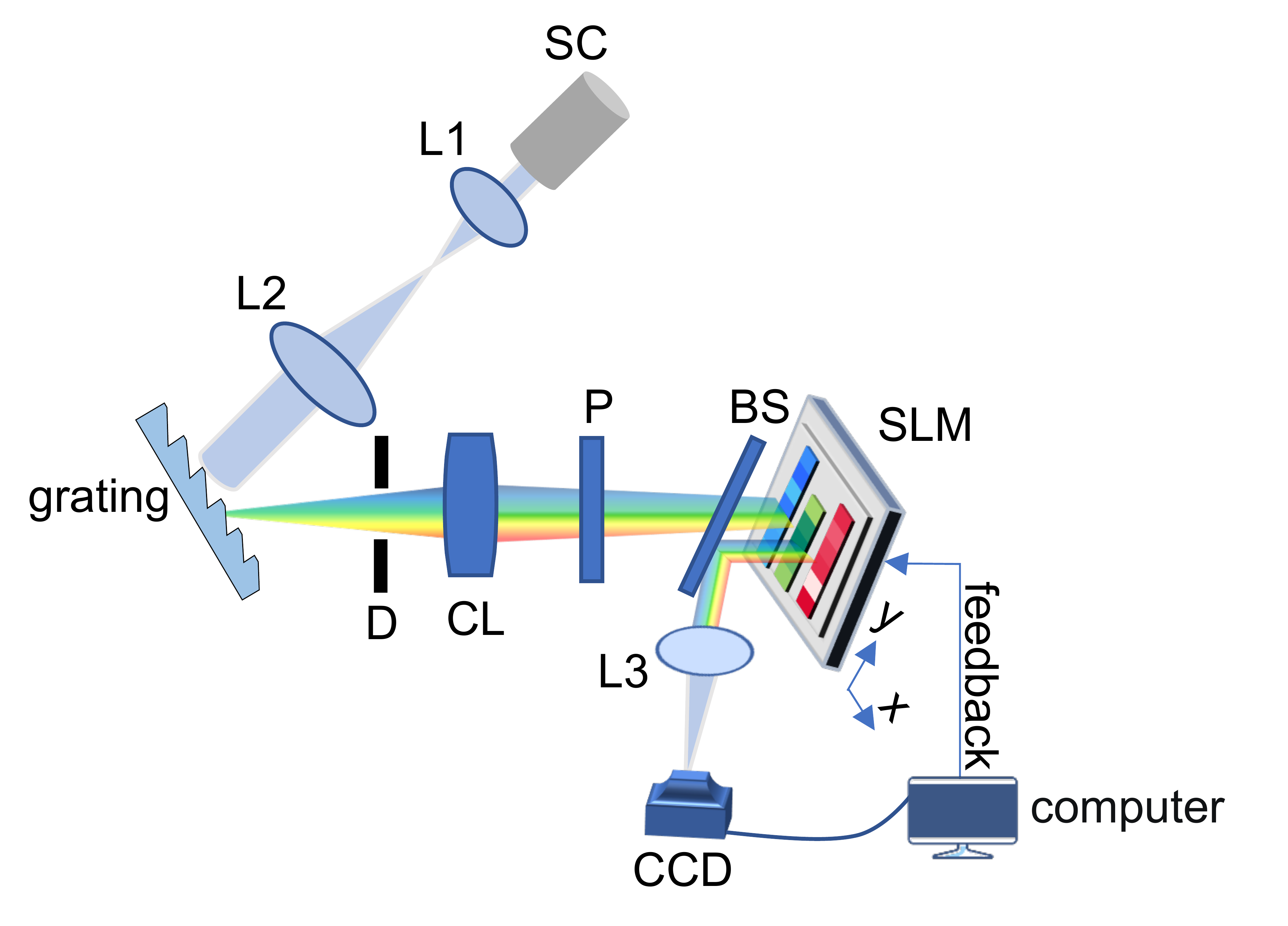}}
\caption{\label{fig:S2}Experimental setup of the wavelength-division multiplexing optical Ising simulator. SC: super-continuum laser; L1, L2 and L3: Fourier lens; D: diaphragm; CL: cylindrical lens; P: polarizer; BS: beam splitter; SLM: spatial light modulator; CCD: charge-coupled device.}
\end{figure}

Figure~\ref{fig:S2} illustrates the experimental setup of the wavelength-division multiplexing optical Ising simulator. Here a super-continuum laser (Anyang SC-5) is used to generate a collimated Gaussian beam. The waist radius of the light beam is enlarged tenfold by the lens ${{L}_{1}}$ (focal length is 50mm) and ${{L}_{2}}$ (focal length is 500mm). The light then illuminates a reflective diffraction grating (inscribed line density of 600/mm), and the cylindrical lens (CL, the focal length is 100mm) focuses the light with different wavelengths onto the SLM (Holoeye PLUTO-NIR-011). With this configuration, the light with different wavelengths is diffracted along the $x$ direction of the SLM, while the $y$-directional pixels are coherently illuminated by the collimated incident light with the same wavelength. An adjustable diaphragm is designed to change the wavelength range incident on the SLM. For the $\pm J$ model and Sherrington-Kirkpatrick (SK) models (Figure~2 in the main text) 80 spins cover wavelengths from 588nm to 611nm, with each spin occupying $\Delta \lambda=0.3$nm in the $x$ direction. For the ${{J}_{1}}\texttt{-}{{J}_{2}}$ model (Figs.~3-4 in the main text), we carry out experiments with 64 spins and set $\kappa _j^k=0$ and $\sigma_j=1$ for the other 16 spins. According to Eq.~(\ref{eq:S6}), $\kappa _j^k=0$ results in no contribution of $\sigma_j$ to $I_k(0,0)$. Polarizer P is used to make the incident beam linearly polarized along the long display axis of the SLM.

Lens ${{L}_{3}}$ (200mm focal length) performs a Fourier transformation of the optical field from the SLM. A CCD (Ophir SP620) is placed at the back focus plane to detect optical field intensity. Due to the finite size of SLM pixel and the resolution of CCD, we integrate the intensity within a region around the center point instead of measuring the intensity at a single point. The effective squared detection area $A$ is defined as $|{u}_{x}|,|{u}_{y}|<\frac{d}{2}$. We find that the results are convergent and stable when $d=45{{\mu}}$m, which corresponds to a detection area covering $10 \times 10$ pixels.

\end{document}